\documentclass[acs, jctc, amsmath,amssymb, reprint]{revtex4-1}

\usepackage{svg}
\usepackage{todonotes}

\usepackage{dsfont}
\usepackage{lipsum}
\usepackage{xcolor}
\usepackage{enumitem}
\usepackage{graphicx}% Include figure files
\usepackage{dcolumn}% Align table columns on decimal point
\usepackage{bm}% bold math
\usepackage{gensymb}
\usepackage{bbm}
\usepackage{todonotes}
\usepackage{soul}
\usepackage[table]{xcolor}

\begin{document}
\clearpage

% \title{ Active learning of machine-learned interatomic potentials using transition path sampling enables the discovery of reaction mechanisms} 
\title{Discovering Reaction Mechanisms with Transition Path Sampling-Based Active Learning of Machine-Learned Potentials}

\author{Ashique Lal }
\affiliation{Van 't Hoff Institute for Molecular Sciences, Universiteit van Amsterdam, Science Park 904, 1098 XH Amsterdam, The Netherlands}
\author{Rik S. Breebaart}
\affiliation{Van 't Hoff Institute for Molecular Sciences, Universiteit van Amsterdam, Science Park 904, 1098 XH Amsterdam, The Netherlands}

\author{Peter G. Bolhuis}
\affiliation{Van 't Hoff Institute for Molecular Sciences, Universiteit van Amsterdam, Science Park 904, 1098 XH Amsterdam, The Netherlands}
\author{Evert Jan Meijer}
\affiliation{Van 't Hoff Institute for Molecular Sciences, Universiteit van Amsterdam, Science Park 904, 1098 XH Amsterdam, The Netherlands}
\email{e.j.meijer@uva.nl}


\begin{abstract}
Machine-learned interatomic potentials (MLPs) provide near density functional theory (DFT) accuracy at reduced computational cost, but their reliability depends on representative training data and often deteriorates in transition-state regions governing rare events. We introduce an active-learning framework in which Transition Path Sampling (TPS) serves as a targeted data-generation engine for constructing MLPs accurate in barrier regions. TPS generates ensembles of unbiased reactive trajectories, and a committee-based uncertainty estimate identifies configurations for selective DFT labeling and retraining. Iterating this cycle systematically refines the potential energy surface in dynamically relevant regions, without the need of prior knowledge of the mechanism. Applied to electrochemical CO$_2$ reduction to CO on copper in explicit water, the approach removes nonphysical artifacts present in early models, achieves near-DFT energy and force accuracy, and enables stable long-time sampling of reactive pathways. Extended TPS simulations reveal multiple dynamically accessible protonation mechanisms. This work establishes TPS as an efficient and principled active-learning strategy for reactive molecular simulations at electrochemical interfaces.
\end{abstract}

\maketitle

\section{Introduction}
Understanding activated processes at electrochemical interfaces is central to heterogeneous catalysis, energy conversion, and electrochemistry. Electrochemical reactions such as CO$_2$ reduction and hydrogen evolution at metal-water interface involve the collective interplay of bond rearrangements, proton transfer, mass transport and solvent reorganization, and surface polarization, all of which evolve on disparate time and length scales. While ab initio molecular dynamics (AIMD) based on density functional theory (DFT) provides an accurate description of these effects, its computational cost severely limits accessible simulation times, rendering direct observation of rare reactive events impractical \cite{Izvekov2001,chen_accelerating_2023}.

Machine-learned interatomic potentials (MLPs) offer a promising route to overcome these limitations by reproducing DFT-level energies and forces at a fraction of the computational cost. Modern architectures based on equivariant message passing have demonstrated remarkable accuracy and stability across a wide range of materials and chemical environments \cite{Batatia2022MACE}. However, the predictive reliability of an MLP is fundamentally limited by the representativeness of its training data. Configurations obtained from equilibrium molecular dynamics predominantly sample harmonic fluctuations around stable basins, while the transition-state regions governing reaction kinetics remain essentially unsampled. As a result, even highly accurate MLPs may exhibit large and uncontrolled errors precisely in the regions most relevant for chemical reactivity \cite{Kulichenko2024Review}.

To improve data efficiency, many workflows employ active learning in which simulations with a provisional potential are iteratively combined with uncertainty-based selection of new configurations for ab initio labeling and retraining (e.g., concurrent learning frameworks) \cite{Zhang2020DPGEN,Kulichenko2024Review,Jinnouchi2019,kulichenko_uncertaintydriven_2023}. 
For reactive or rare-event problems, however, purely unbiased equilibrium exploration often fail to generate transition-region configurations within feasible simulation time, causing uncertainty-driven selection to stall \cite{Fedik2025CaOMLPTPS}. 
Accordingly, a substantial body of recent work couples active learning to biased or enhanced sampling (e.g. well-tempered metadynamics, steered/constrained dynamics, or uncertainty-driven biasing) to force exploration of barrier regions and collect reactive training data \cite{Vitartas2025ALMetaD,Jung2024ALRareEvents,kulichenko_uncertaintydriven_2023,Perego2024DEAL,David2024ArcaNN}. 
While these strategies efficiently discover high-energy configurations, the resulting data are generally generated under an explicit bias potential or external driving protocol, and must be interpreted accordingly \cite{zhang_combining_2024a}.
In contrast, our approach uses transition path sampling to generate an ensemble of unbiased reactive trajectories (with respect to the current surrogate dynamics), and leverages committee uncertainty to target ab initio labeling specifically where the transition-path ensemble reveals insufficient model support.
This combination positions TPS as an active-learning engine tailored to activated dynamics, simultaneously refining the potential in the barrier region and yielding statistically meaningful pathway information without the need of prior knowledge of the exact mechanisms that govern the transitions.

Transition Path Sampling (TPS) provides a fundamentally different approach by sampling ensembles of reactive trajectories rather than configurations \cite{Dellago1998,Bolhuis2002}. In TPS, Monte Carlo moves in trajectory space generate unbiased transition paths connecting metastable states without imposing external bias on the underlying dynamics and without requiring prior knowledge of a reaction coordinate. The resulting transition-path ensemble captures transition-state regions, mechanistic heterogeneity, and dynamical correlations that are inaccessible to equilibrium or biased sampling methods \cite{zhang_combining_2024a}. While early examples of DFT-based TPS \cite{Geissler2001,Ensing2002,Basner2005,Knott2013,Tiwari2016,Dzierlenga2016,Moqadam2017,Moqadam2018,Leitold2020,Paul2020} yielded great mechanistic insight,
despite its conceptual advantages, the high computational cost of DFT when generating large numbers of trajectories has historically restricted TPS to low-dimensional model systems or  models that employ empirical force fields. 

The combination of TPS with machine-learned potentials provides a natural opportunity to overcome these limitations \cite{David2024, Fedik2025CaOMLPTPS}. MLPs enable efficient generation of reactive trajectories, while TPS ensures targeted exploration of the barrier region that dominates reaction kinetics. In this work, we go beyond using MLPs merely as an acceleration tool for TPS and instead employ TPS as an \emph{active-learning engine} for constructing machine-learned potentials that are accurate specifically in the transition region. The procedure starts with a MLP foundation model on which TPS can be performed. If the foundation model is not sufficiently adequate one can also start with an initial dataset consisting of stable-state configurations and approximate barrier-crossing structures obtained from constrained simulations, then train an initial MLP capable of sustaining reactive dynamics. TPS performed on this approximate potential then generates a diverse ensemble of reactive trajectories that repeatedly traverse the barrier region. 

To systematically improve the model, we exploit uncertainty estimates derived from the MLP (by training a committee of MLP models) to identify configurations along TPS trajectories where the potential exhibits poor reliability. These configurations are selectively recomputed at the DFT level and incorporated into the training set, yielding an iterative learning cycle: TPS sampling on the current model, uncertainty-driven selection of informative configurations, \emph{ab initio} labeling, and retraining. Repeating this procedure progressively refines the potential energy surface in the dynamically relevant regions, while simultaneously enabling broader exploration beyond the limited pathways sampled through constrained dynamics.

We apply this framework to the electrochemical reduction of CO$_2$ to CO on a copper slab in explicit water, a system in which proton transfer and interfacial solvent fluctuations play a central mechanistic role. The results presented here employ the MACE equivariant message-passing architecture \cite{Batatia2022MACE}, although the proposed TPS-driven training strategy is independent of the specific MLP formulation. By unifying unbiased rare-event sampling with uncertainty-guided learning, the present approach provides a principled route toward constructing machine-learned potentials capable of describing electrochemical reaction dynamics with near-ab initio fidelity at tractable computational cost and simultaneously allow for the investigation of reaction coordinates and may reveal new unknown pathways otherwise not discovered when employing a biased sampling approach.
\begin{figure}[ht]
\centering
  \includegraphics[width=0.9\linewidth]{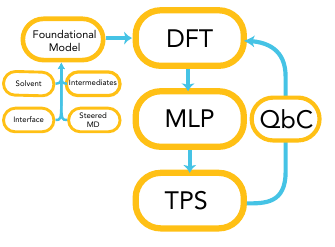}
  \caption{Active learning workflow using MLP foundation models or pretrained model with DFT data. This model can be used as the underlying potential governing the dynamics in TPS simulations to gather data and reactive paths, these trajectories can be analyzed using Query by Committee uncertainty to determine new selected configurations to perform DFT calculations on (gathering new labeled data in regions previously unexplored or where model uncertainty is high). This leads to an improved MLP model which can again be used for TPS, repeating this cycle.}
  \label{fig:workflow_schematic}
 \end{figure}

\section{Methods}

The methodological framework combines TPS, MLPs, and uncertainty-driven active learning into a single iterative loop. The workflow proceeds as follows: (i) an initial MLP is trained on configurations representative of stable states and approximate transition structures; (ii) TPS is performed using this surrogate potential to generate an ensemble of reactive trajectories; (iii) committee-based uncertainty estimates are evaluated along these trajectories to identify configurations where the model is unreliable; (iv) selected configurations are recomputed at the DFT level and incorporated into retraining. This cycle is repeated until the reactive region is consistently described with low uncertainty. A schematic overview is provided in Figure~\ref{fig:workflow_schematic}.

As a representative test case, we consider the elementary protonation–dissociation step in electrochemical CO\textsubscript{2} reduction to CO at a Cu–water interface. The reaction involves formation of a COOH intermediate followed by C–O bond cleavage, a process strongly coupled to interfacial solvent structure and proton mobility \cite{alsunni_electrocatalytic_2021,sheng_electrochemical_2017a}. The collective solvent reorganization and interfacial polarization make this system particularly challenging for both rare-event sampling and machine-learned potential construction.  Therefore, such a realistic condensed-phase reactive environment, provides a great test case for evaluating the methodological framework developed here.

\subsection{Electronic-structure reference calculations}
High-level electronic-structure calculations provide the reference data underpinning the active-learning procedure. All reference energies and forces used for training and validation were computed using DFT as implemented in CP2K software package, which serves as the ground-truth description of bond breaking and formation in this work. Computational details (exchange–correlation functional, basis sets, pseudopotentials, and numerical parameters) are provided in the Supporting Information. DFT calculations are performed only for selected configurations generated during the iterative learning procedure. This starts with 2000 configurations in the first generation and 1000 are added for each concurrent generation. This is the equivalent to 1 ps of DFT MD sampling with a timestep of 1 fs.

\subsection{Machine-learned potentials}
Machine-learned interatomic potentials approximate the potential energy surface by learning the mapping from atomic configurations to energies and forces using the reference electronic-structure data. We employ the Message Passing Atomic Cluster Expansion (MACE) framework \cite{Batatia2022MACE}, which combines a systematically improvable atomic cluster expansion with equivariant message passing to achieve high accuracy and favorable computational scaling (see SI for details on the specific MACE architecture and training settings). The protocol introduced here is, however, not restricted to MACE and can be applied to any MLP architecture capable of providing energies and forces. 

\subsection{Transition Path Sampling}
TPS is employed to sample reactive trajectories connecting predefined metastable states $A$ and $B$. New trajectories are proposed via two way shooting moves and accepted obeying detailed balance \cite{Dellago1998, Bolhuis2002}, yielding an unbiased ensemble of transition paths without requiring a predefined reaction coordinate. This makes TPS particularly suitable for interfacial reactions involving collective solvent and surface rearrangements, where multiple mechanisms may coexist and identifying a single low-dimensional reaction coordinate is nontrivial. Details of the shooting protocol are provided in the Supporting Information.

\begin{figure}[h]
\centering
  \includegraphics[width=8cm,page=4]{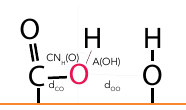}
  \caption{A schematic representation of the system descriptors. $CN_H(O)$ gives a continuous measure of the number of hydrogens attached to a given oxygen (which is indicated in red). $d_{co}$ indicates the distance between the carbon and oxygen undergoing protonation. The distance $d_{OO}$ measures the separation between the oxygen atom in the COOH intermediate and the hydroxyl species that was generated after the initial protonation step. $A(OH)$ is defined as the angle between the OH bond vector ($O\rightarrow H$) and the normal to the plane of the metal surface (indicated as a horizontal orange line).}
  \label{fig1.2}
 \end{figure}

We focus on the reaction from
the unprotonated adsorbed CO\textsubscript{2} configuration (state $A$) and the protonated, dissociated product state yielding CO (state $B$). The boundary conditions defining these states are expressed in terms of physically interpretable collective variables (CVs) which are schematically illustrated in Fig.~\ref{fig1.2}.
The first CV is a smooth measure of the hydrogen coordination number of the oxygen atoms in CO\textsubscript{2}, $CN_H(O)$, defined as
\begin{equation}
    CN_H(O) =  \sum_{i\in O} \sum_{j\in H} \frac{1-(r_{ij}/r_o)^N}{1-(r_{ij}/r_o)^D}
\end{equation}
with $N= 12$, $D=24$ and $r_0=1.3$\AA \cite{daub_initio_2019,ilhan_initio_2011}.
The second CV is the C–O bond distance, $d_{CO}$, defined as the maximal instantaneous distance between the carbon atom of adsorbed CO\textsubscript{2} and its oxygen atoms.
These two CVs together define the states with state $A$ defined by $CN_H(O) < 0.2$ and $d_{CO} < 1.3~\text{\AA}$, corresponding to unprotonated CO\textsubscript{2} with an intact C–O bond and state $B$ defined by $CN_H(O) > 0.8$ and $d_{CO} > 3.0~\text{\AA}$, corresponding to a protonated and dissociated configuration in which CO is formed and the C–O bond is cleaved. 
For analysis of the final production TPS simulation, additional CVs are considered. The distance $d_{OO}$ measures the separation between the oxygen atom in the COOH intermediate and the hydroxyl species generated during the initial protonation step. The angular descriptor $A(OH)$ is defined as the angle between the OH bond vector ($O\rightarrow H$) and the normal to the metal surface (Fig.~\ref{fig1.2}). These variables enable mechanistic discrimination between distinct protonation pathways and provide structural insight beyond the boundary-defining order parameters.

\subsection{Iterating training framework}

 \begin{figure*}[t]
\centering
  \includegraphics[width=\linewidth]{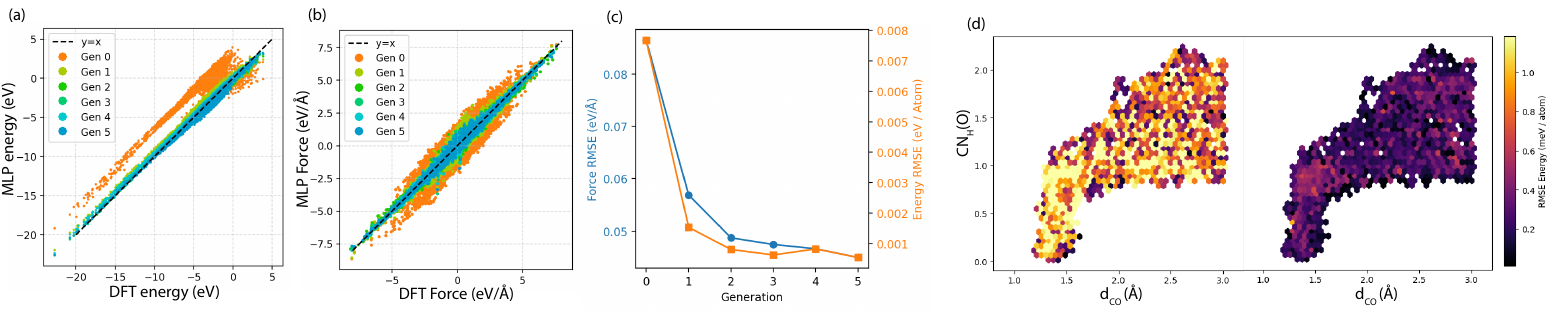}
  \includegraphics[width=\linewidth,page=2]{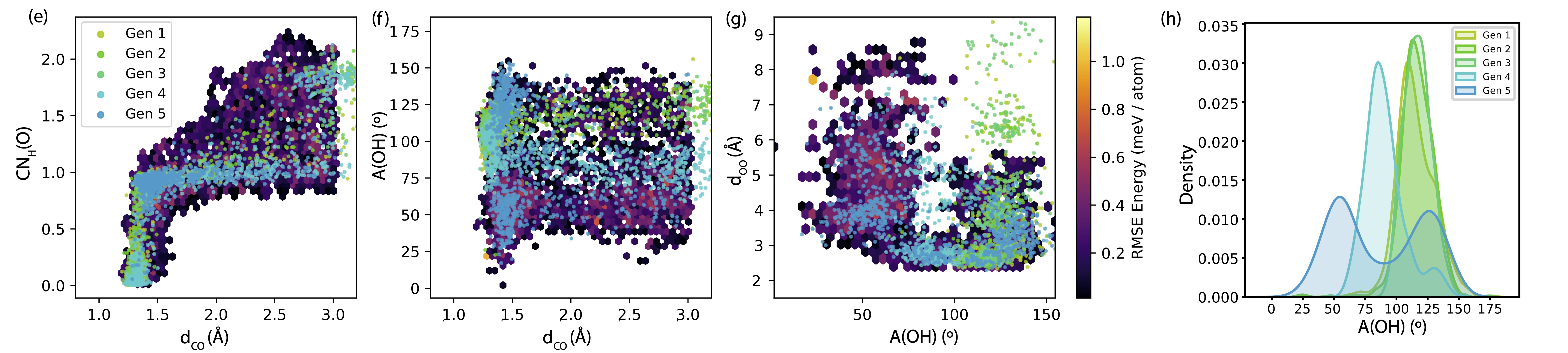}
  \caption{   Parity plot of the DFT target and MLP prediction (a) energies and (b) forces.
(c) RMSE between the DFT and MLP predictions of the energy per atom and forces as a function of generation. 
  (d) RMSE projected onto selected collective variables, for the first generation and 5th generation MLP
  |
  |
  (e-g) Configurations selected for retraining using the committee variance added during each generation on top of the Gen 5 model RMSE (relative to DFT energies) against selected collective variables. 
   (h)Distribution of the training configurations added during each generation projected onto $A(OH)$. }
  \label{fig2}
 \end{figure*}

The workflow is continued iteratively. An initial MLP, trained on DFT energies and forces of 2000 configurations representative of stable states and approximate barrier-crossing structures through a steered MD generated using a foundation MACE model (details of the initial model generation can be found in the Supplementary Information), is first used to generate a reactive trajectory. The TPS is then performed entirely on the MLP potential to produce an ensemble of barrier-crossing trajectories.
These reactive trajectories are extended by 0.5 ps on both sides to provide additional sampling of the stable-state regions.

Multiple independently initialized MLPs are trained to form a committee. The variance across committee predictions along the extended TPS trajectories, as well as along rejected paths, is used to identify high-uncertainty configurations. High-uncertainty regions are selected based on the variance of three observables: total energy, force vectors, and force vectors localized on CO$_2$. This focuses on the selection of regions of interest. To ensure comparable contributions, the variances are normalized across the dataset
\begin{equation}
R_{k,i} = \frac{\sigma_{k,i}^2}{\sum_j \sigma_{k,j}^2},
\end{equation}
where $\sigma_{k,i}^2$ denotes the committee variance of observable $k$ for configuration $i$. The normalized contributions of the variance for the three observables are combined into an uncertainty score
\begin{equation}
S_i = \sum_{k=1}^{3} R_{k,i}.
\end{equation}
Configurations are then selected via importance sampling with probability
\begin{equation}
w_i =
\frac{\lambda S_i + (1-\lambda)}
{\sum_j \left[\lambda S_j + (1-\lambda)\right]},
\qquad \lambda \in [0,1],
\end{equation}
where $\lambda$ controls the balance between uncertainty-driven selection and uniform sampling, ensuring that low-variance configurations remain sampled to improve training stability \cite{wilson_batch_2022}. In our case, $\lambda=0.7$ is used, which balances the addition of training points focused on the high variance regions with more uniformly sampled configurations.

The selected configurations are evaluated at the DFT level, added to the training set, and the MLP is retrained. TPS simulations are subsequently repeated with the updated model. This cycle continues until the committee variance stabilizes and no new high-uncertainty regions are detected.

\begin{figure*}[ht]
\centering
  \includegraphics[width=\linewidth]{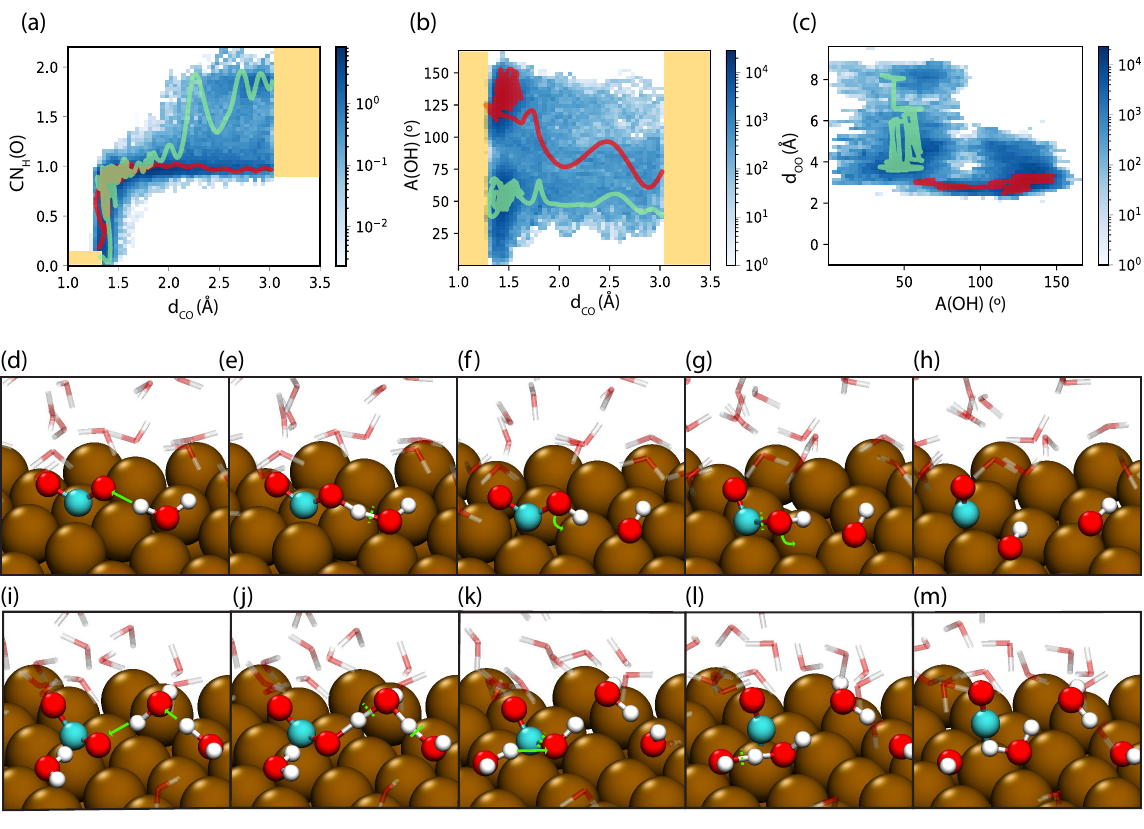}
  \caption{
(a--c) Path histograms projected onto selected collective variables. Two representative trajectories are highlighted: pathway~1 (red) and pathway~2 (green). In pathway~1, the oxygen atom of CO$_2$ that is protonated through a neighboring water molecule remains in an OH configuration during the transition from COOH to CO. During this process the intermediate rotates such that the oxygen approaches the copper surface. In contrast, pathway~2 involves proton transfer through the surrounding water network, allowing the OH species to exchange with neighboring water molecules before CO formation.
(d--h) Representative snapshots illustrating the sequence of events along pathway~1. Green arrows indicate molecular rotations or bond formation, while dashed lines denote bond breaking.
(i--m) Representative snapshots illustrating pathway~2.
}

\label{fig:tps_pathway}
\end{figure*}
 
\section{Results and Discussion}
We first examine the evolution of the machine-learned potential across successive active-learning generations, assessing both quantitative accuracy and qualitative stability of the sampled dynamics on the transition. We then analyze the reactive trajectory ensemble obtained with the converged model, focusing on mechanistic features and pathway diversity revealed by long TPS simulations of the  electrochemical $\rm CO_2$-reduction reaction.

\subsection{Evolution of the Model}

Five successive models (Gen 0 -- Gen 5) were constructed using the iterative active-learning protocol.
To quantify the improvement across generations, we evaluate the root-mean-square error (RMSE) and parity between MLP predictions and DFT reference energies and forces on a set of unseen samples across the path ensembles obtained through TPS using the Gen 5 model.
Figures~\ref{fig2}(a,b) display parity plots comparing MLP and DFT energies and forces. While the initial generation (Gen0) exhibits noticeable deviations from the diagonal, successive generations show progressively tighter agreement. By Gen 3 -- Gen 5, the predictions closely align with the DFT reference values, demonstrating systematic improvement of the potential across the active-learning iterations.
As shown in Fig.~\ref{fig2}(c), the energy per atom RMSE decreases below $1\times10^{-3}$~eV atom$^{-1}$, and the force RMSE falls below $5\times10^{-2}$~eV~\AA$^{-1}$ by Gen 2 -- Gen 3. Beyond Gen 3, both metrics exhibit only minor changes, indicating that the model has reached numerical convergence. 

In parallel with this quantitative convergence, the nature of the sampled configurations changes across generations. Because no reaction coordinate is imposed, TPS generates transition paths determined solely by the dynamics.
As the TPS sampling proceeds, the paths progressively broadens in collective variable space from Gen 0 to Gen 5, visiting regions that where previously unexplored by the model (Fig.~\ref{fig2}e-g) and sampling an entirely new mechanism. Early generations sample a relatively narrow region, whereas later generations populate previously unexplored areas(Fig.~\ref{fig2}h). The configurations selected for retraining increasingly originate from these newly visited regions, indicating that the active learning loop systematically expands the explored configurational space optimizing the potential there.

In Gen0, TPS trajectories contain several nonphysical configurations (see snapshots in SI1), such as distorted interfacial structures and unrealistic hydrogen-bond networks. These configurations correspond to high committee variance and are automatically added to the Gen 1 training set. An effect of the non-physical behavior can be seen in Fig.~\ref{fig2}(f) where configurations with high OH separation that are not present in the subsequent Gen 5 ensemble are selected as training points for Gen 1. In subsequent generations, such artifacts are largely absent, showing that the model learns to eliminate nonphysical regions of phase space. 
Starting from Gen 3, the path ensemble begins to populate a distinct region of collective variable space that is not accessed in earlier generations (Fig.~\ref{fig2}h). This appears as a separated cluster distinguished by the OH angle to the surface normal in the projected histograms(Fig.~\ref{fig2}f-g), consistent with the onset of a new reaction pathway in which the OH coming from the protonation is pointing towards the copper surface . The emergence of this additional mechanism is not enforced but arises from the iterative coupling between TPS sampling and uncertainty-based selection. By Gen 4–5, trajectories repeatedly visit this new region, indicating that the model has reached sufficient local accuracy to sustain stable sampling of multiple pathways. The mechanistic differences between these pathways are analyzed in the following section.

\subsection{Discovery of new pathways}
To assess the stability and exploratory capacity of the final model and to gather mechanistic understanding of the process of CO\textsubscript{2} reduction, a long TPS simulation was performed using Gen 5 for 5000 Monte Carlo two-way shooting steps, yielding an average acceptance rate of $\approx 11\%$. In addition to stable sampling, this extended run revealed the emergence of a mechanistically distinct reaction pathway.

In the initial state, activated CO\textsubscript{2} adsorbed on the surface exhibits asymmetric oxygens bonded to the carbon, with one O atom closer to the surface than the other. This asymmetry allows protonation to occur at two inequivalent sites.

The first reactive pathway begins with protonation of the oxygen farther from the surface by a surface-bound water molecule, forming a COOH intermediate (Fig.~\ref{fig:tps_pathway}d-f). In this configuration, the newly formed OH group initially points toward the surface ($A(OH)<90\degree$) , where the hydrogen atom is closer to the surface than the oxygen ($O\rightarrow H$ aligned opposite to the surface normal). This OH group remains relatively isolated from the surface water network since the oxygen is away from the surface. Therefore, during C–O bond cleavage to form CO, the OH group also rotates to get the O closer to the surface(Fig.~\ref{fig:tps_pathway}f-h). Thus, this mechanism proceeds through an initial proton transfer and subsequent reorientation of the OH group during the C–O bond breaking. 

After approximately 2200 MC steps, TPS transitions to a different pathway. In this alternative mechanism, the oxygen atom closer to the surface gets protonated to form an OH group that points upward ($A(OH)<90\degree$), where the oxygen atom is closer to the surface than ($O\rightarrow H$ aligned parallel to the surface normal).To form such an OH group the oxygen atom has to be protonated by a water molecule that is not adsorbed on the surface, which in turn recieves a proton from a surface bound water molecule (Fig.~\ref{fig:tps_pathway}i-k). Thus the proton has to "hop" between water molecules to form the product surface-bound OH group.  Furthermore, the oxygen atom in this OH group is close to the surface and forms hydrogen bonds with nearby surface bound water molecules. This configuration enables a second proton transfer through the hydrogen-bond network on the surface, allowing the OH species to once again "hop" between neighboring water molecules (Fig.~\ref{fig:tps_pathway}k-m).

The distinction between the two pathways is evident when projecting the path ensemble onto $A(OH)$ vs.\ $d_{CO}$ (Figure~\ref{fig:tps_pathway}b). Two separated peaks appear, corresponding to different OH orientations at similar C–OH bond distances. Further separation is observed in the $A(OH)$ vs.\ $d_{OO}$ histogram
Pathway~1 exhibits shorter $d_{OO}$ distances, indicating closer proximity between the COOH intermediate and the byproduct OH (Figure~\ref{fig:tps_pathway}c). In contrast, Pathway~2 shows larger $d_{OO}$ values, indicating that C–O bond cleavage proceeds without requiring proximity to the hydroxyl byproduct.

Interestingly, the histogram peaks corresponding to the two pathways exhibit similar intensity, and the associated transition paths have comparable lengths (mean trajectory lengths of pathway~1 $\sim620$fs and pathway~2 $\sim640$fs). While more extensive TPS simulations would be required to obtain reliable quantitative branching ratios, this observation indicates that both mechanisms are dynamically accessible within the sampled transition path ensemble generated with the final MLP model and proceed on similar timescales.

This observation is notable in the broader context of CO\textsubscript{2} reduction, where interfacial hydroxyl groups have frequently been implicated in modulating reactivity \cite{liu_unveiling_2025,zhang_regulated_2023a}. The present results suggest that hydroxyl proximity can influence not only stabilization of intermediates but also the mechanistic sequence of protonation and bond cleavage events. 

\section{Conclusion}

We have presented a TPS-driven active-learning framework for constructing machine-learned interatomic potentials that are accurate in transition-state regions governing activated interfacial reactions. By using TPS not only as a rare-event sampling method but also as a targeted data-generation engine, the approach iteratively refines the potential energy surface where kinetics are determined. Committee-based uncertainty along reactive trajectories identifies configurations for selective DFT labeling, enabling systematic improvement of the model while limiting the number of expensive reference calculations.

Applied to electrochemical CO$_2$ reduction to CO on Cu surface in explicit water, the iterative MLP–TPS loop removes nonphysical artifacts present in early surrogate models, achieves near-DFT accuracy in energies and forces, and enables stable long-time sampling of reactive pathways. Extended TPS simulations reveal multiple dynamically accessible protonation mechanisms, demonstrating that the framework supports mechanistic discovery inaccessible to biased methods in addition to improving potential fidelity.

By unifying unbiased transition-path sampling with uncertainty-guided learning, TPS becomes a principled and efficient active-learning strategy for modeling rare reactive events. The framework is broadly applicable to complex condensed-phase reactions where accurate description of barrier regions is essential for predictive simulation. 

\section{Acknowledgements}
AL was supported by the Advanced Research Center for Chemical Building Blocks, ARC CBBC, which is co-founded and co-financed by the Netherlands Organisation for Scientific Research (NWO) and the Netherlands Ministry of Economic Affairs.

\bibliographystyle{achemso}
\bibliography{bibliography}

@article{Dellago1998,
  author = {Dellago, Christoph and Bolhuis, Peter G. and Csajka, F. S. and Chandler, David},
  title = {Transition path sampling and the calculation of rate constants},
  journal = {Journal of Chemical Physics},
  volume = {108},
  pages = {1964--1977},
  year = {1998}
}

@article{Bolhuis2002,
  author = {Bolhuis, Peter G. and Chandler, David and Dellago, Christoph and Geissler, Phillip L.},
  title = {Transition path sampling: Throwing ropes over rough mountain passes, in the dark},
  journal = {Annual Review of Physical Chemistry},
  volume = {53},
  pages = {291--318},
  year = {2002}
}

@article{Batatia2022MACE,
  author = {Batatia, Ilyes and Kov{\'a}cs, D{\'a}vid P. and Simm, Gregor N. C. and Ortner, Christoph and Cs{\'a}nyi, G{\'a}bor},
  title = {MACE: Higher Order Equivariant Message Passing Neural Networks for Fast and Accurate Force Fields},
  journal = {arXiv preprint arXiv:2206.07697},
  year = {2022}
}

@article{Jinnouchi2019,
  title = {On-the-Fly Machine Learning Force Field Generation: {{Application}} to Melting Points},
  shorttitle = {On-the-Fly Machine Learning Force Field Generation},
  author = {Jinnouchi, Ryosuke and Karsai, Ferenc and Kresse, Georg},
  date = {2019-07-17},
  year = 2019,
  journal = {Physical Review B},
  shortjournal = {Phys. Rev. B},
  volume = {100},
  number = {1},
  pages = {014105},
  publisher = {American Physical Society},
  doi = {10.1103/PhysRevB.100.014105},
  url = {https://link.aps.org/doi/10.1103/PhysRevB.100.014105},
  urldate = {2026-03-03},
}

@article{chen_accelerating_2023,
  title = {Accelerating Explicit Solvent Models of Heterogeneous Catalysts with Machine Learning Interatomic Potentials},
  author = {Chen, Benjamin W. J. and Zhang, Xinglong and Zhang, Jia},
  year = 2023,
  journal = {Chemical Science},
  volume = {14},
  number = {31},
  pages = {8338--8354},
  publisher = {The Royal Society of Chemistry},
  doi = {10.1039/D3SC02482B},
}

@article{Izvekov2001,
  title = {Ab Initio Molecular Dynamics Simulation of the {{Ag}}(111)-Water Interface},
  author = {Izvekov, Sergei and Voth, Gregory A.},
  year = 2001,
  month = oct,
  journal = {The Journal of Chemical Physics},
  volume = {115},
  number = {15},
  eprint = {https://pubs.aip.org/aip/jcp/article-pdf/115/15/7196/19302712/7196\_1\_online.pdf},
  pages = {7196--7206},
  issn = {0021-9606},
  doi = {10.1063/1.1403438},
}

@article{Zhang2020DPGEN,
  title = {{{DP-GEN}}: {{A}} Concurrent Learning Platform for the Generation of Reliable Deep Learning Based Potential Energy Models},
  author = {Zhang, Yuzhi and Wang, Haidi and Chen, Weijie and Zeng, Jinzhe and Zhang, Linfeng and Wang, Han and E, Weinan},
  year = 2020,
  journal = {Computer Physics Communications},
  volume = {253},
  pages = {107206},
  issn = {0010-4655},
  doi = {10.1016/j.cpc.2020.107206},
}

@article{Kulichenko2024Review,
  title = {Data Generation for Machine Learning Interatomic Potentials and Beyond},
  author = {Kulichenko, Maksim and Nebgen, Benjamin and Lubbers, Nicholas and Smith, Justin S. and Barros, Kipton and Allen, Alice E. A. and Habib, Adela and Shinkle, Emily and Fedik, Nikita and Li, Ying Wai and Messerly, Richard A. and Tretiak, Sergei},
  year = {2024},
  journal = {Chemical Reviews},
  volume = {124},
  number = {24},
  eprint = {https://doi.org/10.1021/acs.chemrev.4c00572},
  pages = {13681--13714},
  doi = {10.1021/acs.chemrev.4c00572}
}

@article{kulichenko_uncertaintydriven_2023,
  title = {Uncertainty-Driven Dynamics for Active Learning of Interatomic Potentials},
  author = {Kulichenko, Maksim and Barros, Kipton and Lubbers, Nicholas and Li, Ying Wai and Messerly, Richard and Tretiak, Sergei and Smith, Justin S. and Nebgen, Benjamin},
  date = {2023-03},
  year = 2023,
  journaltitle = {Nature Computational Science},
  shortjournal = {Nat Comput Sci},
  volume = {3},
  number = {3},
  pages = {230--239},
  publisher = {Nature Publishing Group},
  issn = {2662-8457},
  doi = {10.1038/s43588-023-00406-5},
  url = {https://www.nature.com/articles/s43588-023-00406-5},
  urldate = {2026-03-03},
  langid = {english}
}

@article{David2024ArcaNN,
  title = {{{ArcaNN}}: Automated Enhanced Sampling Generation of Training Sets for Chemically Reactive Machine Learning Interatomic Potentials},
  shorttitle = {{{ArcaNN}}},
  author = {David, Rolf and Puente, Miguel and Gomez, Axel and Anton, Olaia and Stirnemann, Guillaume and Laage, Damien},
  date = {2025-01-15},
  year = 2025,
  journal = {Digital Discovery},
  shortjournal = {Digital Discovery},
  volume = {4},
  number = {1},
  pages = {54--72},
  publisher = {RSC},
  issn = {2635-098X},
  doi = {10.1039/D4DD00209A},
  url = {https://pubs.rsc.org/en/content/articlelanding/2025/dd/d4dd00209a},
  urldate = {2026-03-03},
  langid = {english}
}

@article{Jung2024ALRareEvents,
  title = {Active Learning of Neural Network Potentials for Rare Events},
  author = {Jung, Gang Seob and Choi, Jong Youl and Lee, Sangkeun Matthew},
  date = {2024-03-13},
  year = 2024,
  journal = {Digital Discovery},
  shortjournal = {Digital Discovery},
  volume = {3},
  number = {3},
  pages = {514--527},
  publisher = {RSC},
  issn = {2635-098X},
  doi = {10.1039/D3DD00216K},
  url = {https://pubs.rsc.org/en/content/articlelanding/2024/dd/d3dd00216k},
  urldate = {2026-03-03},
  langid = {english}
}

@article{Perego2024DEAL,
  title = {Data Efficient Machine Learning Potentials for Modeling Catalytic Reactivity via Active Learning and Enhanced Sampling},
  author = {Perego, Simone and Bonati, Luigi},
  date = {2024-12-19},
  year = 2024,
  journal = {npj Computational Materials},
  shortjournal = {npj Comput Mater},
  volume = {10},
  number = {1},
  pages = {291},
  publisher = {Nature Publishing Group},
  issn = {2057-3960},
  doi = {10.1038/s41524-024-01481-6},
  url = {https://www.nature.com/articles/s41524-024-01481-6},
  urldate = {2026-03-03},
  langid = {english}
}

@article{Vitartas2025ALMetaD,
  title = {Active Learning Meets Metadynamics: Automated Workflow for Reactive Machine Learning Interatomic Potentials},
  shorttitle = {Active Learning Meets Metadynamics},
  author = {Vitartas, Valdas and Zhang, Hanwen and Juraskova, Veronika and Johnston-Wood, Tristan and Duarte, Fernanda},
  date = {2026-01-21},
  yeaer = 2026,
  journal = {Digital Discovery},
  shortjournal = {Digital Discovery},
  volume = {5},
  number = {1},
  pages = {108--122},
  publisher = {RSC},
  issn = {2635-098X},
  doi = {10.1039/D5DD00261C},
  url = {https://pubs.rsc.org/en/content/articlelanding/2026/dd/d5dd00261c},
  urldate = {2026-03-03},
  langid = {english}
}

@article{Fedik2025CaOMLPTPS,
  title = {Challenges and Opportunities for Machine Learning Potentials in Transition Path Sampling: Alanine Dipeptide and Azobenzene Studies},
  shorttitle = {Challenges and Opportunities for Machine Learning Potentials in Transition Path Sampling},
  author = {Fedik, Nikita and Li, Wei and Lubbers, Nicholas and Nebgen, Benjamin and Tretiak, Sergei and Li, Ying Wai},
  year = 2025,
  month = may,
  journal = {Digital Discovery},
  volume = {4},
  number = {5},
  pages = {1158--1175},
  publisher = {RSC},
  issn = {2635-098X},
  doi = {10.1039/D4DD00265B},
  urldate = {2026-03-03},
  langid = {english},
  }

@article{zhang_combining_2024a,
  title = {Combining {{Transition Path Sampling}} with {{Data-Driven Collective Variables}} through a {{Reactivity-Biased Shooting Algorithm}}},
  author = {Zhang, Jintu and Zhang, Odin and Bonati, Luigi and Hou, TingJun},
  year = 2024,
  month = jun,
  journal = {Journal of Chemical Theory and Computation},
  volume = {20},
  number = {11},
  pages = {4523--4532},
  publisher = {American Chemical Society},
  issn = {1549-9618},
  doi = {10.1021/acs.jctc.4c00423},
  urldate = {2026-03-03},
}

@article{liu_unveiling_2025,
  title = {Unveiling Co-Acting Effects of Potassium and Hydroxide Ions on Carbon Dioxide Reduction Reaction Selectivity},
  author = {Liu, Lin and Jiao, Dongxu and Jin, Zhaoyong and Lu, Wenting and Dong, Yilong and Ding, Shuai and Duan, Luotian and Yao, Mingguang and Xu, Shan and Liu, Yanhua and Zhang, Lei and Fan, Jinchang and Cui, Xiaoqiang},
  year = 2025,
  month = jun,
  journal = {Journal of Colloid and Interface Science},
  volume = {688},
  pages = {591--599},
  issn = {0021-9797},
  doi = {10.1016/j.jcis.2025.02.184},
  urldate = {2026-03-03},
}

@article{zhang_regulated_2023a,
  title = {Regulated {{CO}} Adsorption by the Electrode with {{OH}}- Repulsive Property for Enhancing {{C}}--{{C}} Coupling},
  author = {Zhang, Qixing and Ren, Dan and Gao, Jing and Wang, Zhongke and Wang, Juan and Pan, Sanjiang and Wang, Manjing and Luo, Jingshan and Zhao, Ying and Gr{\"a}tzel, Michael and Zhang, Xiaodan},
  year = 2023,
  month = sep,
  journal = {Green Chemical Engineering},
  volume = {4},
  number = {3},
  pages = {331--337},
  issn = {26669528},
  doi = {10.1016/j.gce.2022.07.007},
  urldate = {2026-03-03},
  langid = {english}
}

@article{alsunni_electrocatalytic_2021,
  title = {Electrocatalytic {{Reduction}} of {{CO2}} to {{CO}} over {{Ag}}(110) and {{Cu}}(211) {{Modeled}} by {{Grand-Canonical Density Functional Theory}}},
  author = {Alsunni, Yousef A. and Alherz, Abdulaziz W. and Musgrave, Charles B.},
  year = 2021,
  month = nov,
  journal = {The Journal of Physical Chemistry C},
  volume = {125},
  number = {43},
  pages = {23773--23783},
  publisher = {American Chemical Society},
  issn = {1932-7447},
  doi = {10.1021/acs.jpcc.1c07484},
  urldate = {2026-03-03}
}

@article{sheng_electrochemical_2017a,
  title = {Electrochemical Reduction of {{CO}} 2 into {{CO}} on {{Cu}}(100): A New Insight into the {{C}}--{{O}} Bond Breaking Mechanism},
  shorttitle = {Electrochemical Reduction of {{CO}} 2 into {{CO}} on {{Cu}}(100)},
  author = {Sheng, Tian and Sun, Shi-Gang},
  year = 2017,
  journal = {Chemical Communications},
  volume = {53},
  number = {17},
  pages = {2594--2597},
  publisher = {Royal Society of Chemistry},
  doi = {10.1039/C6CC08583K},
  urldate = {2026-03-03},
  langid = {english}
}

@article{daub_initio_2019,
  title = {Ab {{Initio Molecular Dynamics Simulations}} of the {{Influence}} of {{Lithium Bromide Salt}} on the {{Deprotonation}} of {{Formic Acid}} in {{Aqueous Solution}}},
  author = {Daub, Christopher D. and Halonen, Lauri},
  year = 2019,
  month = aug,
  journal = {The Journal of Physical Chemistry B},
  volume = {123},
  number = {31},
  pages = {6823--6829},
  publisher = {American Chemical Society},
  issn = {1520-6106},
  doi = {10.1021/acs.jpcb.9b04618},
  urldate = {2026-03-03}
}

@article{ilhan_initio_2011,
  title = {Ab Initio Molecular Dynamics of Proton Networks in Narrow Polymer Electrolyte Pores},
  author = {Ilhan, Mehmet A and Spohr, Eckhard},
  year = 2011,
  month = may,
  journal = {Journal of Physics: Condensed Matter},
  volume = {23},
  number = {23},
  pages = {234104},
  issn = {0953-8984},
  doi = {10.1088/0953-8984/23/23/234104},
  urldate = {2026-03-03},
  langid = {english}
}

@article{wilson_batch_2022,
  title = {Batch Active Learning for Accelerating the Development of Interatomic Potentials},
  author = {Wilson, Nathan and Willhelm, Daniel and Qian, Xiaoning and Arr{\'o}yave, Raymundo and Qian, Xiaofeng},
  year = 2022,
  month = jun,
  journal = {Computational Materials Science},
  volume = {208},
  pages = {111330},
  issn = {0927-0256},
  doi = {10.1016/j.commatsci.2022.111330},
  urldate = {2026-03-05}
}

@misc{batatia2025foundationmodelatomisticmaterials,
      title={A foundation model for atomistic materials chemistry}, 
      author={Ilyes Batatia and Philipp Benner and Yuan Chiang and Alin M. Elena and Dávid P. Kovács and Janosh Riebesell and Xavier R. Advincula and Mark Asta and Matthew Avaylon and William J. Baldwin and Fabian Berger and Noam Bernstein and Arghya Bhowmik and Filippo Bigi and Samuel M. Blau and Vlad Cărare and Michele Ceriotti and Sanggyu Chong and James P. Darby and Sandip De and Flaviano Della Pia and Volker L. Deringer and Rokas Elijošius and Zakariya El-Machachi and Fabio Falcioni and Edvin Fako and Andrea C. Ferrari and John L. A. Gardner and Mikolaj J. Gawkowski and Annalena Genreith-Schriever and Janine George and Rhys E. A. Goodall and Jonas Grandel and Clare P. Grey and Petr Grigorev and Shuang Han and Will Handley and Hendrik H. Heenen and Kersti Hermansson and Christian Holm and Cheuk Hin Ho and Stephan Hofmann and Jad Jaafar and Konstantin S. Jakob and Hyunwook Jung and Venkat Kapil and Aaron D. Kaplan and Nima Karimitari and James R. Kermode and Panagiotis Kourtis and Namu Kroupa and Jolla Kullgren and Matthew C. Kuner and Domantas Kuryla and Guoda Liepuoniute and Chen Lin and Johannes T. Margraf and Ioan-Bogdan Magdău and Angelos Michaelides and J. Harry Moore and Aakash A. Naik and Samuel P. Niblett and Sam Walton Norwood and Niamh O'Neill and Christoph Ortner and Kristin A. Persson and Karsten Reuter and Andrew S. Rosen and Louise A. M. Rosset and Lars L. Schaaf and Christoph Schran and Benjamin X. Shi and Eric Sivonxay and Tamás K. Stenczel and Viktor Svahn and Christopher Sutton and Thomas D. Swinburne and Jules Tilly and Cas van der Oord and Santiago Vargas and Eszter Varga-Umbrich and Tejs Vegge and Martin Vondrák and Yangshuai Wang and William C. Witt and Thomas Wolf and Fabian Zills and Gábor Csányi},
      year={2025},
      eprint={2401.00096},
      archivePrefix={arXiv},
      primaryClass={physics.chem-ph},
      url={https://arxiv.org/abs/2401.00096}, 
}

@article{mermin_thermal_1965,
  title = {Thermal {{Properties}} of the {{Inhomogeneous Electron Gas}}},
  author = {Mermin, N. David},
  year = 1965,
  month = mar,
  journal = {Phys. Rev.},
  volume = {137},
  number = {5A},
  pages = {A1441-A1443},
  issn = {0031-899X},
  doi = {10.1103/PhysRev.137.A1441},
  urldate = {2026-01-04},
  copyright = {http://link.aps.org/licenses/aps-default-license},
  langid = {english}
}

@article{goedecker_separable_1996,
  title = {Separable Dual-Space {{Gaussian}} Pseudopotentials},
  author = {Goedecker, S. and Teter, M. and Hutter, J.},
  year = 1996,
  month = jul,
  journal = {Phys. Rev. B},
  volume = {54},
  number = {3},
  pages = {1703--1710},
  publisher = {American Physical Society},
  doi = {10.1103/PhysRevB.54.1703},
  urldate = {2024-06-28}
}

@article{grimme_consistent_2010,
  title = {A Consistent and Accurate Ab Initio Parametrization of Density Functional Dispersion Correction ({{DFT-D}}) for the 94 Elements {{H-Pu}}},
  author = {Grimme, Stefan and Antony, Jens and Ehrlich, Stephan and Krieg, Helge},
  year = 2010,
  month = apr,
  journal = {J. Chem. Phys.},
  volume = {132},
  number = {15},
  pages = {154104},
  issn = {0021-9606},
  doi = {10.1063/1.3382344},
  urldate = {2025-12-09}
}

@article{perdew_generalized_1996,
  title = {Generalized {{Gradient Approximation Made Simple}}},
  author = {Perdew, John P. and Burke, Kieron and Ernzerhof, Matthias},
  year = 1996,
  month = oct,
  journal = {Phys. Rev. Lett.},
  volume = {77},
  number = {18},
  pages = {3865--3868},
  publisher = {American Physical Society},
  doi = {10.1103/PhysRevLett.77.3865},
  urldate = {2025-12-09}
}

@article{_cp2k_,
  title = {{{CP2K}}: {{An}} Electronic Structure and Molecular Dynamics Software Package - {{Quickstep}}: {{Efficient}} and Accurate Electronic Structure Calculations},
  shorttitle = {{{CP2K}}},
  author = {Kühne, Thomas D. and Iannuzzi, Marcella and Del Ben, Mauro and Rybkin, Vladimir V. and Seewald, Patrick and Stein, Frederick and Laino, Teodoro and Khaliullin, Rustam Z. and Schütt, Ole and Schiffmann, Florian and Golze, Dorothea and Wilhelm, Jan and Chulkov, Sergey and Bani-Hashemian, Mohammad Hossein and Weber, Valéry and Borštnik, Urban and Taillefumier, Mathieu and Jakobovits, Alice Shoshana and Lazzaro, Alfio and Pabst, Hans and Müller, Tiziano and Schade, Robert and Guidon, Manuel and Andermatt, Samuel and Holmberg, Nico and Schenter, Gregory K. and Hehn, Anna and Bussy, Augustin and Belleflamme, Fabian and Tabacchi, Gloria and Glöß, Andreas and Lass, Michael and Bethune, Iain and Mundy, Christopher J. and Plessl, Christian and Watkins, Matt and VandeVondele, Joost and Krack, Matthias and Hutter, Jürg},
  date = {2020-05-19},
  year = 2020,
  journaltitle = {The Journal of Chemical Physics},
  shortjournal = {J. Chem. Phys.},
  volume = {152},
  number = {19},
  pages = {194103},
  issn = {0021-9606},
  doi = {10.1063/5.0007045},
  url = {https://doi.org/10.1063/5.0007045},
  urldate = {2026-04-02}
}

@article{kohn_selfconsistent_1965,
  title = {Self-{{Consistent Equations Including Exchange}} and {{Correlation Effects}}},
  author = {Kohn, W. and Sham, L. J.},
  year = 1965,
  month = nov,
  journal = {Phys. Rev.},
  volume = {140},
  number = {4A},
  pages = {A1133-A1138},
  publisher = {American Physical Society},
  doi = {10.1103/PhysRev.140.A1133},
  urldate = {2025-12-09}
}

@article{bonomi_plumed_2009,
  title = {{{PLUMED}}: {{A}} Portable Plugin for Free-Energy Calculations with Molecular Dynamics},
  shorttitle = {{{PLUMED}}},
  author = {Bonomi, Massimiliano and Branduardi, Davide and Bussi, Giovanni and Camilloni, Carlo and Provasi, Davide and Raiteri, Paolo and Donadio, Davide and Marinelli, Fabrizio and Pietrucci, Fabio and Broglia, Ricardo A. and Parrinello, Michele},
  date = {2009-10-01},
  year = 2009,
  journaltitle = {Computer Physics Communications},
  shortjournal = {Computer Physics Communications},
  volume = {180},
  number = {10},
  pages = {1961--1972},
  issn = {0010-4655},
  doi = {10.1016/j.cpc.2009.05.011},
  url = {https://www.sciencedirect.com/science/article/pii/S001046550900157X},
  urldate = {2026-03-06}
}

@article{bonomi_promoting_2019,
  title = {Promoting Transparency and Reproducibility in Enhanced Molecular Simulations},
  author = {Bonomi, Massimiliano and Bussi, Giovanni and Camilloni, Carlo and Tribello, Gareth A. and Banáš, Pavel and Barducci, Alessandro and Bernetti, Mattia and Bolhuis, Peter G. and Bottaro, Sandro and Branduardi, Davide and Capelli, Riccardo and Carloni, Paolo and Ceriotti, Michele and Cesari, Andrea and Chen, Haochuan and Chen, Wei and Colizzi, Francesco and De, Sandip and De La Pierre, Marco and Donadio, Davide and Drobot, Viktor and Ensing, Bernd and Ferguson, Andrew L. and Filizola, Marta and Fraser, James S. and Fu, Haohao and Gasparotto, Piero and Gervasio, Francesco Luigi and Giberti, Federico and Gil-Ley, Alejandro and Giorgino, Toni and Heller, Gabriella T. and Hocky, Glen M. and Iannuzzi, Marcella and Invernizzi, Michele and Jelfs, Kim E. and Jussupow, Alexander and Kirilin, Evgeny and Laio, Alessandro and Limongelli, Vittorio and Lindorff-Larsen, Kresten and Löhr, Thomas and Marinelli, Fabrizio and Martin-Samos, Layla and Masetti, Matteo and Meyer, Ralf and Michaelides, Angelos and Molteni, Carla and Morishita, Tetsuya and Nava, Marco and Paissoni, Cristina and Papaleo, Elena and Parrinello, Michele and Pfaendtner, Jim and Piaggi, Pablo and Piccini, GiovanniMaria and Pietropaolo, Adriana and Pietrucci, Fabio and Pipolo, Silvio and Provasi, Davide and Quigley, David and Raiteri, Paolo and Raniolo, Stefano and Rydzewski, Jakub and Salvalaglio, Matteo and Sosso, Gabriele Cesare and Spiwok, Vojtěch and Šponer, Jiří and Swenson, David W. H. and Tiwary, Pratyush and Valsson, Omar and Vendruscolo, Michele and Voth, Gregory A. and White, Andrew and {The PLUMED consortium}},
  date = {2019-08},
  year = 2019,
  journaltitle = {Nature Methods},
  shortjournal = {Nat Methods},
  volume = {16},
  number = {8},
  pages = {670--673},
  publisher = {Nature Publishing Group},
  issn = {1548-7105},
  doi = {10.1038/s41592-019-0506-8},
  url = {https://www.nature.com/articles/s41592-019-0506-8},
  urldate = {2026-03-06},
  langid = {english}
}

@article{eastman_openmm_2024,
  title = {{{OpenMM}} 8: {{Molecular Dynamics Simulation}} with {{Machine Learning Potentials}}},
  shorttitle = {{{OpenMM}} 8},
  author = {Eastman, Peter and Galvelis, Raimondas and Peláez, Raúl P. and Abreu, Charlles R. A. and Farr, Stephen E. and Gallicchio, Emilio and Gorenko, Anton and Henry, Michael M. and Hu, Frank and Huang, Jing and Krämer, Andreas and Michel, Julien and Mitchell, Joshua A. and Pande, Vijay S. and Rodrigues, João PGLM and Rodriguez-Guerra, Jaime and Simmonett, Andrew C. and Singh, Sukrit and Swails, Jason and Turner, Philip and Wang, Yuanqing and Zhang, Ivy and Chodera, John D. and De Fabritiis, Gianni and Markland, Thomas E.},
  date = {2024-01-11},
  year = 2024,
  journaltitle = {The Journal of Physical Chemistry B},
  shortjournal = {J. Phys. Chem. B},
  volume = {128},
  number = {1},
  pages = {109--116},
  publisher = {American Chemical Society},
  issn = {1520-6106},
  doi = {10.1021/acs.jpcb.3c06662},
  url = {https://doi.org/10.1021/acs.jpcb.3c06662},
  urldate = {2026-03-06}
}

@article{swenson_openpathsampling_2019,
  title = {{{OpenPathSampling}}: {{A Python Framework}} for {{Path Sampling Simulations}}. 1. {{Basics}}},
  shorttitle = {{{OpenPathSampling}}},
  author = {Swenson, David W. H. and Prinz, Jan-Hendrik and Noe, Frank and Chodera, John D. and Bolhuis, Peter G.},
  date = {2019-02-12},
  year = 2019,
  journaltitle = {Journal of Chemical Theory and Computation},
  shortjournal = {J. Chem. Theory Comput.},
  volume = {15},
  number = {2},
  pages = {813--836},
  publisher = {American Chemical Society},
  issn = {1549-9618},
  doi = {10.1021/acs.jctc.8b00626},
  url = {https://doi.org/10.1021/acs.jctc.8b00626},
  urldate = {2026-03-06}
}

@article{tribello_plumed_2014,
  title = {{{PLUMED}} 2: {{New}} Feathers for an Old Bird},
  shorttitle = {{{PLUMED}} 2},
  author = {Tribello, Gareth A. and Bonomi, Massimiliano and Branduardi, Davide and Camilloni, Carlo and Bussi, Giovanni},
  date = {2014-02-01},
  year = 2014,
  journaltitle = {Computer Physics Communications},
  shortjournal = {Computer Physics Communications},
  volume = {185},
  number = {2},
  pages = {604--613},
  issn = {0010-4655},
  doi = {10.1016/j.cpc.2013.09.018},
  url = {https://www.sciencedirect.com/science/article/pii/S0010465513003196},
  urldate = {2026-03-06}
}

@article{David2024,
  title = {Competing Reaction Mechanisms of Peptide Bond Formation in Water Revealed by Deep Potential Molecular Dynamics and Path Sampling},
  volume = {146},
  ISSN = {1520-5126},
  url = {http://dx.doi.org/10.1021/jacs.4c03445},
  DOI = {10.1021/jacs.4c03445},
  number = {20},
  journal = {Journal of the American Chemical Society},
  publisher = {American Chemical Society (ACS)},
  author = {David,  Rolf and Tu\~n\'on,  I{\~}naki and Laage,  Damien},
  year = 2024,
  month = may,
  pages = {14213–14224}
}

@article{Geissler2001,
  title = {Autoionization in Liquid Water},
  volume = {291},
  ISSN = {1095-9203},
  url = {http://dx.doi.org/10.1126/science.1056991},
  DOI = {10.1126/science.1056991},
  number = {5511},
  journal = {Science},
  publisher = {American Association for the Advancement of Science (AAAS)},
  author = {Geissler,  Phillip L. and Dellago,  Christoph and Chandler,  David and Hutter,  Jürg and Parrinello,  Michele},
  year = {2001},
  month = mar,
  pages = {2121–2124}
}

@article{Ensing2002,
  doi = {10.1021/jp025833l},
  url = {https://doi.org/10.1021/jp025833l},
  year = {2002},
  month = aug,
  publisher = {American Chemical Society ({ACS})},
  volume = {106},
  number = {34},
  pages = {7902--7910},
  author = {Bernd Ensing and Evert Jan Baerends},
  title = {Reaction Path Sampling of the Reaction between Iron({II}) and Hydrogen Peroxide in Aqueous Solution},
  journal = {The Journal of Physical Chemistry A}
}

@article{Tiwari2016,
  doi = {10.1039/c6fd00132g},
  url = {https://doi.org/10.1039/c6fd00132g},
  year = {2016},
  publisher = {Royal Society of Chemistry ({RSC})},
  volume = {195},
  pages = {291--310},
  author = {Ambuj Tiwari and Bernd Ensing},
  title = {Reactive trajectories of the Ru$^{2+/3+}$ self-exchange reaction and the connection to Marcus theory},
  journal = {Faraday Discussions}
}

@article{Moqadam2017,
  doi = {10.1039/c7cp01268c},
  url = {https://doi.org/10.1039/c7cp01268c},
  year = {2017},
  publisher = {Royal Society of Chemistry ({RSC})},
  volume = {19},
  number = {20},
  pages = {13361--13371},
  author = {Mahmoud Moqadam and Enrico Riccardi and Thuat T. Trinh and Anders Lervik and Titus S. van Erp},
  title = {Rare event simulations reveal subtle key steps in aqueous silicate condensation},
  journal = {Physical Chemistry Chemical Physics}
}

@article{Moqadam2018,
  doi = {10.1073/pnas.1714070115},
  url = {https://doi.org/10.1073/pnas.1714070115},
  year = {2018},
  month = apr,
  publisher = {Proceedings of the National Academy of Sciences},
  volume = {115},
  number = {20},
  pages = {E4569--E4576},
  author = {Mahmoud Moqadam and Anders Lervik and Enrico Riccardi and Vishwesh Venkatraman and Bj{\o}rn K{\aa}re Alsberg and Titus S. van Erp},
  title = {Local initiation conditions for water autoionization},
  journal = {Proceedings of the National Academy of Sciences}
}

@article{Basner2005,
  doi = {10.1021/ja043320h},
  url = {https://doi.org/10.1021/ja043320h},
  year = {2005},
  month = oct,
  publisher = {American Chemical Society ({ACS})},
  volume = {127},
  number = {40},
  pages = {13822--13831},
  author = {Jodi E. Basner and Steven D. Schwartz},
  title = {How Enzyme Dynamics Helps Catalyze a Reaction in Atomic Detail:~ A Transition Path Sampling Study},
  journal = {Journal of the American Chemical Society}
}

@article{Knott2013,
  doi = {10.1021/ja410291u},
  url = {https://doi.org/10.1021/ja410291u},
  year = {2013},
  month = dec,
  publisher = {American Chemical Society ({ACS})},
  volume = {136},
  number = {1},
  pages = {321--329},
  author = {Brandon C. Knott and Majid Haddad Momeni and Michael F. Crowley and Lloyd F. Mackenzie and Andreas W. G\"{o}tz and Mats Sandgren and Stephen G. Withers and Jerry St{\aa}hlberg and Gregg T. Beckham},
  title = {The Mechanism of Cellulose Hydrolysis by a Two-Step,  Retaining Cellobiohydrolase Elucidated by Structural and Transition Path Sampling Studies},
  journal = {Journal of the American Chemical Society}
}

@incollection{Dzierlenga2016,
  doi = {10.1016/bs.mie.2016.05.028},
  url = {https://doi.org/10.1016/bs.mie.2016.05.028},
  year = {2016},
  publisher = {Elsevier},
  pages = {21--43},
  author = {M.W. Dzierlenga and M.J. Varga and S.D. Schwartz},
  title = {Path Sampling Methods for Enzymatic Quantum Particle Transfer Reactions},
  booktitle = {Methods in Enzymology}
}

@article{Leitold2020,
  doi = {10.1063/5.0002766},
  url = {https://doi.org/10.1063/5.0002766},
  year = {2020},
  month = jul,
  publisher = {{AIP} Publishing},
  volume = {153},
  number = {2},
  pages = {024103},
  author = {Christian Leitold and Christopher J. Mundy and Marcel D. Baer and Gregory K. Schenter and Baron Peters},
  title = {Solvent reaction coordinate for an {SN}2 reaction},
  journal = {The Journal of Chemical Physics}
}

@article{Paul2020,
  doi = {10.1002/cphc.202000177},
  url = {https://doi.org/10.1002/cphc.202000177},
  year = {2020},
  month = jun,
  publisher = {Wiley},
  volume = {21},
  number = {13},
  pages = {1455--1473},
  author = {Tanmoy Kumar Paul and Srabani Taraphder},
  title = {Coordination Dynamics of Zinc Triggers the Rate Determining Proton Transfer in Human Carbonic Anhydrase {II}},
  journal = {{ChemPhysChem}}
}

@article{vandevondele_gaussian_2007,
  title = {Gaussian Basis Sets for Accurate Calculations on Molecular Systems in Gas and Condensed Phases},
  author = {VandeVondele, Joost and Hutter, Jürg},
  date = {2007-09-18},
  year = 2007,
  journaltitle = {The Journal of Chemical Physics},
  shortjournal = {The Journal of Chemical Physics},
  volume = {127},
  number = {11},
  pages = {114105},
  issn = {0021-9606},
  doi = {10.1063/1.2770708},
  url = {https://doi.org/10.1063/1.2770708},
  urldate = {2024-06-28},}

@article{hjorthlarsen_atomic_2017,
  title = {The Atomic Simulation Environment—a {{Python}} Library for Working with Atoms},
  author = {Hjorth Larsen, Ask and Jørgen Mortensen, Jens and Blomqvist, Jakob and Castelli, Ivano E and Christensen, Rune and Dułak, Marcin and Friis, Jesper and Groves, Michael N and Hammer, Bjørk and Hargus, Cory and Hermes, Eric D and Jennings, Paul C and Bjerre Jensen, Peter and Kermode, James and Kitchin, John R and Leonhard Kolsbjerg, Esben and Kubal, Joseph and Kaasbjerg, Kristen and Lysgaard, Steen and Bergmann Maronsson, Jón and Maxson, Tristan and Olsen, Thomas and Pastewka, Lars and Peterson, Andrew and Rostgaard, Carsten and Schiøtz, Jakob and Schütt, Ole and Strange, Mikkel and Thygesen, Kristian S and Vegge, Tejs and Vilhelmsen, Lasse and Walter, Michael and Zeng, Zhenhua and Jacobsen, Karsten W},
  date = {2017-06},
  year = 2017,
  journaltitle = {Journal of Physics: Condensed Matter},
  shortjournal = {J. Phys.: Condens. Matter},
  volume = {29},
  number = {27},
  pages = {273002},
  publisher = {IOP Publishing},
  issn = {0953-8984},
  doi = {10.1088/1361-648X/aa680e},
  url = {https://doi.org/10.1088/1361-648X/aa680e},
  urldate = {2026-04-15},
  langid = {english}
}
% \printbibliography %Prints bibliography

\end{document}

% \begin{tocentry}

% Some journals require a graphical entry for the Table of Contents.
% This should be laid out ``print ready'' so that the sizing of the

% Inside the \texttt{tocentry} environment, the font used is Helvetica
% 8\,pt, as required by \emph{Journal of the American Chemical

% The surrounding frame is 9\,cm by 3.5\,cm, which is the maximum
% permitted for  \emph{Journal of the American Chemical Society}
% graphical table of content entries. The box will not resize if the
% content is too big: instead it will overflow the edge of the box.

% This box and the associated title will always be printed on a
% separate page at the end of the document.

% \end{tocentry}

\section{Method details}
\subsection{DFT settings}
All electronic structure calculations were performed using density functional theory (DFT) \cite{kohn_selfconsistent_1965}, as implemented in the CP2K software package \cite{_cp2k_}. Electronic exchange-correlation effects were
described using the Perdew-Burke-Ernzerhof (PBE)\cite{perdew_generalized_1996} functional within the
generalized gradient approximation (GGA), complemented by Grimme's
DFT-D3\cite{grimme_consistent_2010} dispersion correction to capture long-range van der Waals
interactions.

CP2K implements the Gaussian and plane-wave (GPW) formalism, combining
localized Gaussian basis functions with an auxiliary plane-wave
expansion of the electronic density. Goedecker-Teter-Hutter (GTH)
pseudopotentials\cite{goedecker_separable_1996} optimized for the PBE functional were employed. A TZV Pbasis set was used for C, H, and O atoms, DZV was adopted for
Cu\cite{vandevondele_gaussian_2007}. The plane-wave cutoff energy was set to 400 Ry. A Fermi-Dirac smearing\cite{mermin_thermal_1965} corresponding to 400K was also applied to improve convergence. 

Cu(100) surfaces were modeled as 4-layer slabs in a 4 × 4
supercell (14.46 × 14.46 Å\textsuperscript{2} for Cu) in a periodic box. The
bottom two layers were fixed to mimic bulk constraints, and a water film
($\approx 13$ Å thick, comprising 98 H\textsubscript{2}O molecules) was placed
above the metal surface, separated by $\approx20$ Å of vacuum to
avoid spurious slab interactions.  All
calculations were performed at the $\Gamma$-point. No extra electrons where added to the system and calculations where done at a constant-charge regime.

\subsection{Initial model setup}

To accelerate the generation of intermediate-state configurations, the MACE-MP-O foundational model \cite{batatia2025foundationmodelatomisticmaterials} was used. This allows decorrelated configurations to be obtained much faster than with a DFT-based protocol. A 1 ns MD simulation at 400 K was run, from which 400 equally spaced configurations were extracted for each of the following molecule adsorbed at the Cu-water interface: *CO\textsubscript{2}, *COOH, and *CO.

In addition, 400 configurations were extracted from a 1 ns simulation of a cubic box (a = 12.43 Å) containing 4 HCOOH, 4 CO, 4 CO\textsubscript{2}, and 32 H\textsubscript{2}O molecules.

To sample configurations along reactive coordinates, steered MD simulations were performed using Atomic Simulation Environment (ASE) \cite{hjorthlarsen_atomic_2017} with PLUMED \cite{bonomi_plumed_2009,tribello_plumed_2014}. First, 200 configurations were obtained from a simulation in which *CO\textsubscript{2} was protonated by the nearest H\textsubscript{2}O molecule, using the distance between the transferring proton and the O atom of* CO\textsubscript{2} as the collective variable. Another 200 configurations were obtained from a similar simulation starting from *COOH (taken from the final configuration of the *COOH MD run), where C–OH bond cleavage to form *CO and *OH was driven using the C–OH bond length as the collective variable.

All simulations employed a Langevin thermostat with a friction coefficient of 1~fs$^{-1}$ and a timestep of 1~fs at a temperature of 400~K. In total, 2000 configurations were generated with this procedure. These configurations were used to train a committee of Gen 0 models. The resulting models are computationally faster than the foundation model and better adapted to the specific configuration space relevant for the transition path sampling simulations performed in this work.
% founddation model used for data generation of 0th genration of sampling and then used to gather dft calculations. Mace model then has later generations lower quality but faster. This due to speed and memory. bla bla bla 
% basically here what we did all details, the training protocol foundation model to start, genreation of stable configurations and a steered md using it, then tos in this initial model (the generaiton of stbale conf etc allows for a committee)

\subsection{MLP training hyperparamter}
A committee of three MACE models was employed where each model was initialized with a different random seed (123, 234, and 345) to form an ensemble. The architecture used a local interaction cutoff radius of 6.0~\AA, 32 channels, a correlation order of 3, and a maximum angular resolution of $L_{\mathrm{max}} = 1$.

The training dataset was constructed iteratively. The initial generation (Gen 0) consisted of 2,000 configurations, and each subsequent generation added 1,000 configurations to the existing dataset.

A 4:1 train–test split was applied, with 10\% of the training set further reserved for validation to monitor convergence. Each model in the committee was trained for 100 epochs using a learning rate of 0.01 and a batch size of 10.

\subsection{TPS Setup}
ASE\cite{hjorthlarsen_atomic_2017} was used as the molecular dynamics engine together with PLUMED\cite{bonomi_plumed_2009,tribello_plumed_2014,bonomi_promoting_2019} to perform steered MD for generating the initial reactive trajectory. Transition path sampling (TPS) was carried out using the OpenPathSampling (OPS \cite{swenson_openpathsampling_2019}) framework with OpenMM\cite{eastman_openmm_2024} as the MD engine through the \texttt{openmmml}\cite{eastman_openmm_2024} python package.

Stable states were defined based on the coordination number ($CN_H(O)$) described in the main text and the C–O bond distance. State A was defined as $CN(\mathrm{O}) < 0.2$ and $d_{CO} < 1.35$~\AA, while State B was defined as $CN(\mathrm{O}) > 0.8$ and $d_{CO} > 3$~\AA. Trajectories were rejected if either of the shooting trajectory length exceeded 2000~fs.

\begin{figure}[h]
\centering
  \includegraphics[page=3,width=\textwidth]{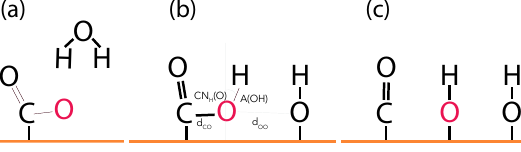}
  \caption{A schematic representation of (a) State A, (b) Transition state and (c) State B. $CN_H(O)$ gives a continuous measure of the number of hydrogens attached to a given oxygen (which is indicated in red). $d_{co}$ indicates the distance between the carbon and oxygen undergoing protonation. The distance $d_{OO}$ measures the separation between the oxygen atom in the *COOH intermediate and the hydroxyl species that was generated after the initial protonation step. $A(OH)$ is defined as the angle between the OH bond vector ($O\rightarrow H$) and the normal to the plane of the metal surface (indicated as the orange line).}\label{fig1.2}
\end{figure}

All simulations employed a Langevin thermostat with a friction coefficient of 1~fs$^{-1}$ and a timestep of 1~fs at a temperature of 400~K. Two-way shooting moves with a uniform selector were used in TPS. For active learning, 2000 Monte Carlo shooting steps were performed per generation, and 5000 MC steps were carried out with the final model for analysis.

\clearpage

\subsection{Evolution of model across generations (Gen 0 -- Gen 2)}{}

 \begin{figure}[h]
\centering
  \includegraphics[width=\linewidth]{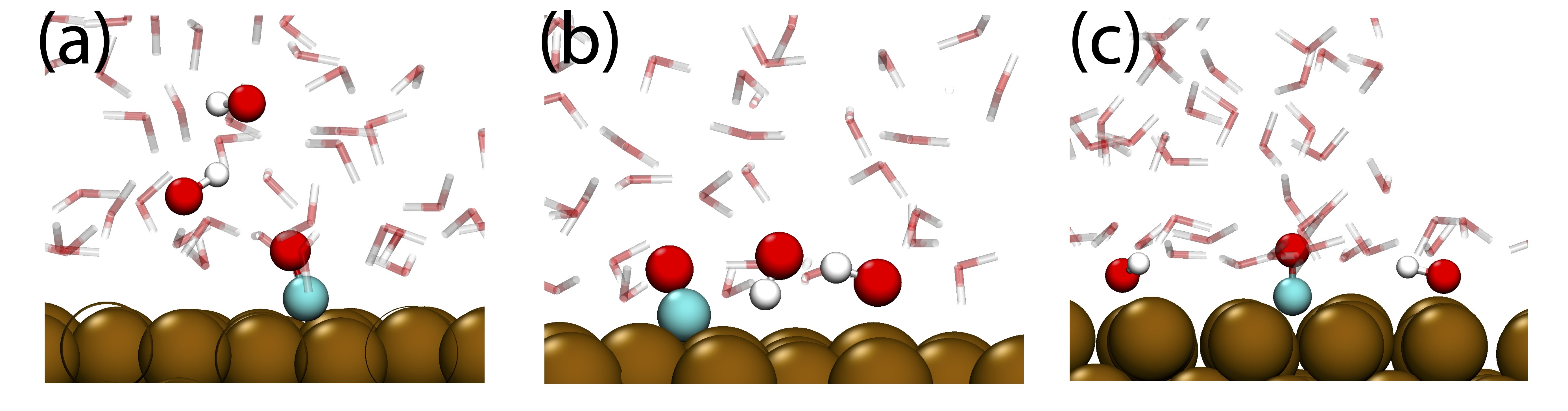}
  \caption{Representative snapshots showing evolution of the unphysical configurations in product-state structure during MD up to Gen 2: (a) Gen 0 : the OH group moves away from the surface into the bulk solvent; (b) Gen 1 : the OH group stays near the surface and is H-bound to a surface site; (c) Gen 2 : the OH group stays near the surface and is O-bound to a surface site.}\label{ch5:a:f:artifacts}
 \end{figure}

\subsection{Proton hopping}

 \begin{figure}[h]
\centering
  \includegraphics[width=\linewidth]{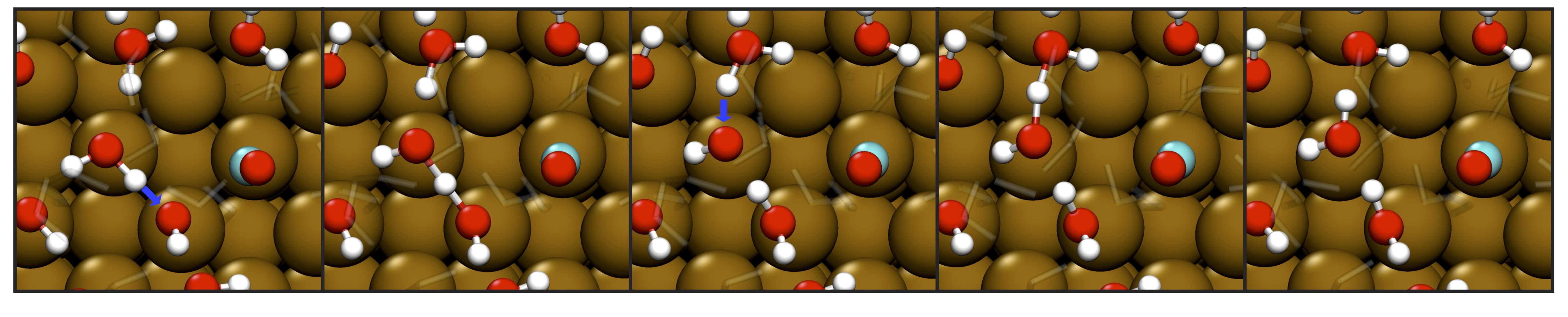}
  \caption{Representative snapshots of proton hopping through 3 different water molecules.}\label{ch5:a:f:proton_hop}
 \end{figure}

 \begin{figure}
 \section{Model improvement}
\centering
  \includegraphics[width=0.75\linewidth]{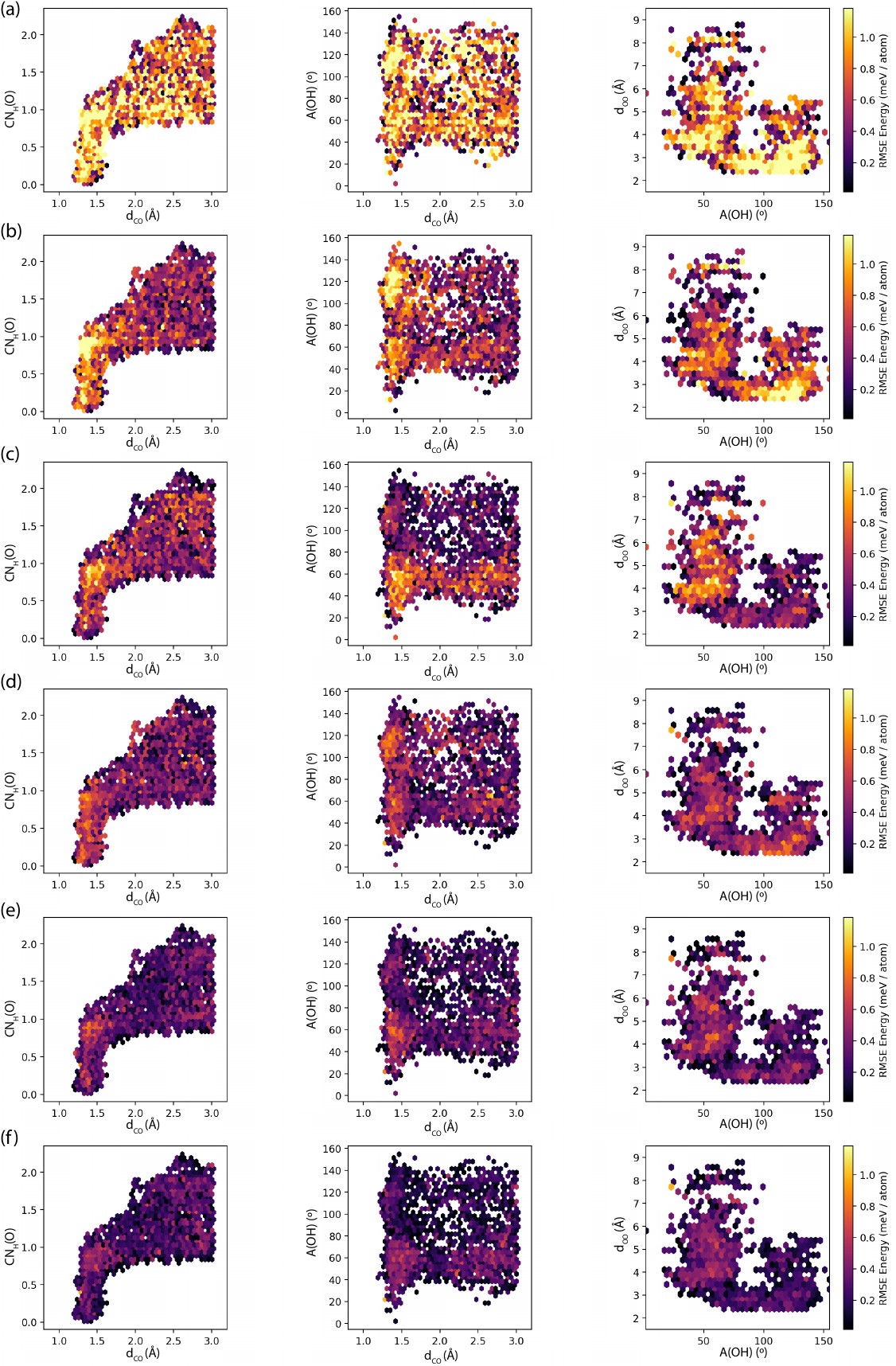}
  \caption{RMSE with respect to DFT projected onto selected collective variables for (a)0th (b) 1st (c) 2nd (d) 3rd (e) 4th (f) 5th Gen. }
  \label{fig:model_improve}
 \end{figure}

\clearpage
% \bibliography{references}
% \bibliographystyle{achemso}
\bibliography{bibliography}
% \printbibliography %Prints bibliography